\documentclass[12pt]{article}
\usepackage{amsmath,amsthm,amscd,amssymb}
\usepackage{latexsym}
\theoremstyle{plain}
\usepackage{epsf}
\usepackage{epsfig}
\newlength{\mycolspace}
\theoremstyle{definition}

\theoremstyle{remark}

\newcommand{\abs}[1]{|#1|}
\newcommand{\norm}[1]{\|#1\|}

\newcommand{\calC}{{\cal{C}}}

\newcommand{\calH}{{\cal{H}}}

\newcommand{\calS}{{\cal{S}}}

\newcommand{\vp}{{\varphi_0}}

\newcommand{\bbC}{{\mathbb{C}}}
\def\re{\mathop{\rm Re}}

\begin{document}
\title{Acceleration-induced nonlocality: kinetic memory versus dynamic memory}
\author{C. Chicone\\Department of Mathematics\\University of 
Missouri-Columbia\\Columbia, Missouri 65211, USA 
\and B. Mashhoon\thanks{Corresponding author. E-mail:
mashhoonb@missouri.edu (B. Mashhoon).\newline Phone: (573) 882-6526;\;
FAX: (573) 882-4195.} \\Department of Physics and
Astronomy\\University of Missouri-Columbia\\Columbia, Missouri 65211, USA}
\maketitle
\begin{abstract} 
The characteristics of the memory of accelerated motion in Min\-kow\-ski 
spacetime are discussed within the framework of 
the nonlocal theory of accelerated observers. 
Two types of memory are distinguished: kinetic 
and dynamic. We show that only kinetic memory is acceptable, since 
dynamic memory leads to divergences for nonuniform accelerated motion.
\end{abstract}
\section{Introduction}

The special theory of relativity is based on two basic postulates: 
Lorentz invariance and the hypothesis of locality. Lorentz invariance 
refers to a fundamental
symmetry principle, namely, the invariance of basic physical laws 
under inhomogeneous Lorentz transformations. In practice these laws 
of nature involve physical
quantities measured by inertial observers in Minkowski spacetime.
An inertial observer always moves uniformly and refers
its observations to the fixed spatial axes
of an inertial frame; it can be depicted by a straight line in the Minkowski 
diagram and represents an
ideal; in fact, physical observers are all effectively accelerated. 
For instance, one can imagine the influence of radiation pressure on 
the path of a cosmic particle.
In general, the acceleration of an observer consists of the translational 
acceleration of its path as well as the rotation of its spatial frame.
Observers with translational acceleration 
are therefore 
represented by curved lines in the Minkowski diagram.
As an example of a rotating observer, consider a uniformly moving
observer that refers its observations to spatial axes that rotate with 
respect to the spatial frame of the underlying inertial coordinate system. 
The hypothesis of locality refers to the
measurements of realistic (i.e. accelerated) observers: such an 
observer is postulated to be equivalent, at each event along its 
worldline, to a momentarily comoving
inertial observer. The origin of this assumption can be traced back 
to the work of Lorentz in the context of his classical electron 
theory~\cite{1}; later, it was
simply adopted as a general rule in relativity theory~\cite{2}.

Along its worldline, the accelerated observer passes through a 
continuous infinity of hypothetical momentarily comoving inertial 
observers. Stated mathematically, the translationally accelerated 
observer's curved worldline
is the \emph{envelope} of the straight worldlines of this class 
of hypothetical inertial observers. Therefore, the
hypothesis of locality has two components: (i) the assumption that 
the measurements of the accelerated observer must be somehow 
connected to the measurements of
the hypothetical class of momentarily comoving inertial observers 
along its worldline and (ii) that this connection is 
postulated to be the pointwise
equivalence of the accelerated observer and the momentarily comoving 
inertial observer. The latter means that the acceleration of the 
observer does not {\it
directly} affect the result of its measurement; devices that obey 
this rule are called ``standard". Thus the hypothesis of locality is 
a simple generalization of
the assumption that the rods and clocks of special relativity theory 
are not directly affected by acceleration
\cite{2}.

What is the physical basis for the hypothesis of locality? It is 
difficult to argue with part (i) of this hypothesis, since the 
fundamental laws of
(nongravitational) physics have been formulated with respect to 
inertial observers and hence the measurements of accelerated 
observers should be in some way
related to those of inertial observers. 
On the other hand, part (ii) can only 
be valid if the measurement process occurs instantaneously and
in a pointwise manner. 
That is, (ii) is appropriate
for phenomena involving coincidences of classical point particles
and null rays. 
Classical waves, 
on the other hand, are extended in time and space with a 
characteristic wave period $T$ and a
corresponding wavelength $\lambda$, respectively. Imagine, for 
example, the measurement of the frequency of an incident 
electromagnetic wave by an accelerated
observer; at least a few periods of the wave must be received by the 
observer before an adequate determination of the frequency would 
become possible. Thus this measurement
process  is nonlocal and extends over the worldline of the observer. 
The observer's acceleration can be characterized by certain 
{\it acceleration lengths} $L$ given by
$c^2/g$ and $c/\Omega$ for translational acceleration $g$ and 
rotational frequency $\Omega$, respectively.
The nonlocality of the external radiation is thus expected to couple
with the intrinsic scales associated with the acceleration of the observer.

Classical wave phenomena are expected to violate the hypothesis of 
locality. The scale of such violation would be given by $\lambda 
/L=T/(L/c)$, where $L/c$ is the
{\it acceleration time}. The hypothesis of locality will hold if 
$\lambda$ is so small that the incident radiation behaves like a ray, 
i.e. in the eikonal (or JWKB)
limit such that
$\lambda /L\to 0$; alternatively, $L$ can be so large that 
deviations of the form $\lambda /L$ would be below the sensitivity 
threshold of the detectors available
at present. Consider, e.g., laboratory experiments on the Earth; 
typical acceleration lengths would be $c^2/g_\oplus \simeq 1\,{\rm lyr}$ and 
$c/\Omega _\oplus \simeq 28\, {\rm AU}$, 
so that for essentially all practical purposes one can ignore 
any possible deviations from locality at the present time. In this 
way, we can account for the
fact that the standard theory of relativity is in agreement with all 
observational data available at present. As a matter of principle, 
however, it is necessary to
contemplate generalizations of the hypothesis of locality in order to 
take due account of intrinsic wave phenomena for realistic 
(accelerated) observers.

All of our considerations in this paper are within the framework of 
classical field theory; nevertheless, it is necessary to remark that 
quantum theory is based on
the notion of wave-particle duality, and so an adequate treatment of 
classical wave phenomena is a necessary prelude to a satisfactory 
quantum theory.

To proceed, we consider the most general extension of the hypothesis 
of locality that is consistent with causality and the superposition 
principle. A nonlocal Lorentz-invariant
theory of accelerated observers has been developed along these
lines~\cite{3,3a, 3b,3c}
and is presented in Section~2. The theory involves a kernel 
that depends primarily on the
acceleration of the observer; that is, the measurements of the 
observer depend on its past history of acceleration. 
The main physical principle that is employed in the nonlocal theory
for the determination of the kernel is the assumption that 
an intrinsic radiation field can never stand completely still with
respect to an accelerated observer;
this statement involves a simple generalization of a property of inertial
observers to all observers.
Thus the 
accelerated observer is endowed with
{\it memory}, and the past affects the present through an averaging 
process, where the weight function is proportional to the kernel 
$K(\tau ,\tau')$. It turns out
that the kernel $K$ cannot be completely determined by the theory 
presented in Section~2. An additional simplifying assumption is 
therefore introduced in Section~3: 
$K(\tau ,\tau ')$ must be a function of a single variable. Two 
cases are then considered:
(1)  $K(\tau ,\tau ')=k_0(\tau ')$ 
and
(2) $K(\tau ,\tau ')=k(\tau-\tau').$ 
We show that case~(1)---i.e. the 
\emph{kinetic} memory case---has acceptable
properties that are described in Section 3. Case~(2), i.e. 
the \emph{dynamic} memory case, is treated in detail in Sections 3 
and 4, where it is shown that the kernel function $k$ can be unbounded
even if the observer's past history has constant velocity except
for one episode of smooth translational acceleration  with finite duration. 
Specifically, we study the measurement of electromagnetic 
radiation fields by an 
observer that undergoes translational or rotational acceleration 
that lasts for only a finite interval of its proper time.
After the acceleration is turned off,
the observer  measures in addition to the regular field 
a residual field that contains the
memory of its past acceleration. This leftover piece is a finite 
\emph{constant} field (\emph{kinetic} memory) in case (1); however,
it is time dependent (\emph{dynamic} memory) in case (2). 
We rule out the latter case, since we prove that the measured field could
diverge under certain reasonable circumstances. We are thus left
with a unique theory that involves kinetic memory. 
An important aspect of our nonlocal ansatz is that the kernel
induced by the acceleration of the observer
depends on the spin of the radiation field under consideration. 
In particular, the kernel vanishes for an intrinsic scalar field,
i.e. such a field is always \emph{local}.
As discussed in Section~5,
our theory therefore rules out the possibility that a pure
scalar (or pseudoscalar) field  exists in nature. 
This conclusion is in agreement with available
experimental data.
The nonlocal theory therefore predicts that any scalar particle would
have to be a composite.
Section 5 contains a  brief discussion and our 
conclusions. 
A detailed discussion of the observational consequences of the nonlocal 
theory is beyond the scope of this work.
In the following, we use units such that
$c=1$, i.e. the speed of light in vacuum is unity.

\section{Accelerated observers and nonlocality}

The measurement of length by accelerated observers involves subtle 
issues in relativity theory that have been investigated in detail 
\cite{4,4a,4b}; for our present
purpose, the main result of such studies is that an accelerated frame 
of reference, i.e. an extended coordinate system set up in the 
neighborhood of an accelerated
observer, is of rather limited theoretical significance. We shall 
therefore refer all measurements to an inertial reference frame in 
Minkowski spacetime.

Imagine a global inertial frame with coordinates $x=(t,\mathbf{ x})$ 
and the standard class of static inertial observers with their 
orthonormal tetrad frame
$\lambda^\mu_{(\alpha)}=\delta ^\mu _\alpha$, where $\lambda^\mu 
_{(0)}$ is the temporal direction at each event and $\lambda^\mu 
_{(i)}$, $i=1,2,3$, are the spatial
directions. The hypothesis of locality implies that an accelerated 
observer is also endowed with a tetrad frame 
$\hat{\lambda}^\mu_{(\alpha)}(\tau )$, where $\tau
$ is the proper time along its worldline. For each $\tau$, 
$\hat{\lambda}^\mu_{(\alpha)}(\tau)$ coincides with the constant 
tetrad frame (related to $\lambda^\mu
_{(\alpha)}$ by a Lorentz transformation) of the momentarily comoving 
inertial observer. We note that 
$
d\hat{\lambda}^\mu_{(\alpha)}/d\tau 
=\phi_\alpha ^{\hspace{0.075in}\beta}\hat{\lambda}^\mu _{(\beta)}
$, 
where $\phi_{\alpha 
\beta}=-\phi_{\beta \alpha}$ is a tensor such that 
$\phi_{0 i}=(\mathbf{g})_i$ and
$\phi_{ij}=\epsilon_{ijk}($\boldmath$\Omega$\unboldmath$)_k$. Here 
$\mathbf{g}(\tau )$ is the translational acceleration of the observer 
and
\boldmath$\Omega$\unboldmath$(\tau )$ is the rotational frequency of 
its spatial frame. 
Each element of the acceleration tensor $\phi_{\alpha\beta}$
is a scalar under the inhomogeneous Lorentz transformations
of the background spacetime.
We assume throughout that the acceleration is 
turned on at $\tau =\tau_0$
and will in general be turned off at $\tau_1>\tau_0$.

Let $f_{\mu \nu}$ represent an electromagnetic radiation field as 
measured by the standard set of static inertial observers. According 
to the hypothesis of
locality $\hat{f}_{\alpha \beta}=f_{\mu \nu}\hat{\lambda 
}^\mu_{(\alpha)}\hat{\lambda}^\nu _{(\beta)}$, i.e. the projection of 
the field on the instantaneous
tetrad frame, would be the field  measured by the 
accelerated observer. On the other hand, let $F_{\alpha \beta }(\tau 
)$ be the true result of such a
measurement. Taking causality into account, the most general linear 
relationship between $F_{\alpha \beta}(\tau )$ and $\hat{f}_{\alpha 
\beta}(\tau )$ is
\begin{equation}\label{eq1} 
F_{\alpha \beta}(\tau )
=\hat{f}_{\alpha \beta}(\tau )
+\int^\tau_{\tau_0}K_{\alpha \beta \gamma\delta }(\tau ,\tau ')
\hat{f}^{\gamma\delta}(\tau')\, d\tau '.
\end{equation}
This relation refers to quantities that are all scalars under
the Poincar\'e group of spacetime transformations of the underlying
inertial coordinate system.
We note that the magnitude of the nonlocal part of equation 
\eqref{eq1} is of the form $\lambda /L$ if the kernel is proportional 
to the acceleration of the
observer. It follows from Volterra's theorem that in the space of 
continuous functions the relationship between $F$ and $f$ is 
unique~\cite{5,5a}; 
this theorem has been
extended to the Hilbert space of square-integrable functions 
by Tricomi~\cite{5b}.

The basic ansatz~\eqref{eq1} is consistent with an observation originally
put forward by Bohr and Rosenfeld that the electromagnetic field cannot
be measured at a spacetime \emph{point}; in fact, an averaging process
is necessary over a spacetime neighborhood~\cite{br,bra}. In the case of
measurements by \emph{inertial} observers envisaged by 
Bohr and Rosenfeld~\cite{br,bra}, there is no intrinsic temporal or 
spatial scale
associated with the inertial observers; 
therefore, one can effectively pass to the
limiting case of a point with no difficulty as
the dimensions of the spacetime neighborhood can be shrunk to zero
without any obstruction. For an accelerated observer, however, the
intrinsic acceleration time and length need to be properly taken into
account. Hence the nonlocal ansatz~\eqref{eq1} may be interpreted in
terms of a certain averaging process over the past worldline of the 
accelerated observer.

To determine the kernel $K$, let us first mention a basic consequence 
of the hypothesis of locality for a radiation field. Imagine plane 
monochromatic
electromagnetic waves of frequency $\omega$ propagating along the 
$z$-axis and an observer rotating uniformly about this axis with 
frequency $\Omega_0$ in the
$(x,y)$-plane on a circle of radius $\rho$ in the underlying
inertial reference frame. We find from 
$\hat{f}_{\alpha \beta}=f_{\mu \nu}\hat{\lambda}^\mu_{(\alpha 
)}\hat{\lambda}^\nu _{(\beta)}$ that
according to the rotating observer the frequency of the wave is 
$\hat{\omega}=\gamma (\omega \mp \Omega_0)$, where $\gamma$ is the 
Lorentz factor corresponding to
the speed $\rho \Omega_0$ of the observer and the upper (lower) sign 
refers to incident positive (negative) helicity radiation. This result 
has a simple intuitive
interpretation: In an incident positive (negative) helicity wave the 
electric and magnetic field vectors rotate with frequency $\omega 
(-\omega)$ about the
direction of propagation of the wave. As seen by the rotating 
observer, the field vectors rotate with frequency 
$\omega-\Omega_0\,(-\omega-\Omega_0)$ with respect to 
the inertial temporal coordinate $t$; moreover, the Lorentz
factor simply accounts for time dilation $dt=\gamma d\tau$. It 
follows that a positive helicity incident wave can stand completely 
still with respect to all observers rotating uniformly
with frequency $\Omega_0=\omega$. In terms of energy, we 
have 
$\hat{E}=\gamma(E-$\boldmath$\sigma$\unboldmath$\cdot$
\boldmath$\Omega$\unboldmath$_0)$, 
where
\boldmath$\sigma$\unboldmath\;  is the spin of the incident photon. More 
generally, for oblique incidence $\hat{E}=\gamma (E- \hbar 
M\Omega_0)$, where $M$ is the
multipole parameter such that $\hbar M$ is the component of the total 
(orbital plus spin) angular momentum along the $z$-axis. 
This is an example of the 
general phenomenon of
spin-rotation coupling; various aspects of this effect and the 
available observational evidence are discussed in~\cite{6,6a,6b,6c,6d,6e}. 
Again, 
the incident wave can
theoretically stand completely still for all observers rotating with 
frequency $\Omega_0$ such that $\omega =M\Omega_0$. Let us recall 
here a fundamental
consequence of Lorentz invariance, namely that a radiation field can 
never stand completely still with respect to an inertial observer. 
That is, an inertial observer can
move along the direction of propagation of a wave so fast that the 
frequency $\hat{\omega}=\gamma \omega (1-\beta)$ can approach zero 
but the mathematical limit of
$\hat{\omega}=0$ is never physically achieved, since the observer's 
speed cannot reach the speed of light in vacuum $(\beta <1)$. 
Therefore, for an inertial observer $\hat{\omega}=0$
implies that $\omega=0$. On the other hand, while we find that the 
hypothesis of locality predicts that a circularly polarized wave can 
stand completely still with
respect to a uniformly rotating observer, this possibility 
can be avoided in the nonlocal theory by an appropriate choice of the 
kernel.

To implement the requirement that a radiation field can never stand 
completely still with respect to any observer, we assume that if 
$F_{\alpha \beta}(\tau )$
turns out to be constant in equation~\eqref{eq1}, then $f_{\mu \nu}$ 
must have been  originally constant just as in the case of
inertial observers in the standard theory of relativity. 
It is convenient to replace the tensor 
$f_{\mu \nu}$ by a six-vector
$f$, with electric and magnetic fields as components, and introduce 
the ``Lorentz'' matrix $\Lambda $ such that $\hat{f}=\Lambda f$. Then for 
constant fields $f$ and $F$,
equation~\eqref{eq1} takes the form
\begin{equation}\label{eq2} F=\Lambda(\tau) f+\int^\tau_{\tau_0}K(\tau 
,\tau ')\Lambda (\tau')f\,d\tau',\end{equation}
where for $\tau=\tau_0$, the matrix 
$\Lambda_0:=\Lambda(\tau_0)$ is constant and
$F=\Lambda_0 f$. Thus in the nonlocal 
theory the kernel $K$ should be determined from the Volterra integral 
equation
\begin{equation}\label{eq3} \Lambda_0=\Lambda (\tau 
)+\int^\tau _{\tau_0}K(\tau ,\tau')\Lambda (\tau ')\,d\tau'.
\end{equation}
It follows from Volterra's theory (see Appendix A) that to every 
kernel $K$ corresponds a unique {\it resolvent} kernel $R(\tau 
,\tau')$ such that
\begin{equation}\label{eq4} \Lambda (\tau )=\Lambda_0
+\int^\tau_{\tau_0}R(\tau ,\tau')\Lambda_0\,d\tau 
'.\end{equation}
Therefore, only the integral of the resolvent kernel is 
completely determined by our physical requirement
\begin{equation}\label{eq5} 
\int^\tau_{\tau_0} R(\tau ,\tau')\,d\tau '
=\Lambda (\tau )\Lambda_0^{-1}-I,\end{equation}
where $I$ is the unit matrix. It is clear at this point that given 
$\Lambda (\tau )$, relations \eqref{eq3}--\eqref{eq5} are not 
sufficient to determine the kernel
$K$ uniquely. To proceed further, other simplifying 
restrictions are necessary on $K$ or $R$,
\begin{equation}\label{eq6} \hat{f}(\tau )
=F(\tau)+\int^\tau_{\tau_0}R(\tau ,\tau')F(\tau ')\,d\tau '.
\end{equation}
This must be done in such a way as to preserve time translation 
invariance in the underlying inertial coordinate system.

Let us finally remark that for a scalar field, $\Lambda (\tau )=1$ and 
equations \eqref{eq3}--\eqref{eq5} simply reduce to the requirement 
that $K(\tau ,\tau')$ must
have a vanishing integral over $\tau':\tau_0\to \tau$. That is, the 
connection between the kernel and the acceleration of the observer 
disappears. This
circumstance is further discussed in Section 5.

\section{Memory}

It is necessary to introduce  simplifying assumptions in order to 
find a unique kernel $K$. We therefore tentatively postulate that $K$ 
is a function of a single
variable. There are two reasonable possibilities:
\begin{align*} \tag{case 1} K(\tau ,\tau')=k_0(\tau ')\\
\intertext{and}
\tag{case 2} K(\tau ,\tau')=k(\tau -\tau');
\end{align*}
in either case, the basic requirement of time translation invariance 
in the background global inertial frame is satisfied.
\subsection{Kinetic memory}
In case~(1), the kernel $k_0$ corresponds to a simple weight 
function that can be determined by differentiating 
equation~\eqref{eq3},
\begin{equation}\label{eq7} k_0(\tau 
)=-\frac{d\Lambda}{d\tau}\Lambda^{-1}(\tau )=\Lambda (\tau 
)\frac{d\Lambda^{-1}}{d\tau }.\end{equation}
The kernel $k_0$ is thus directly proportional to the acceleration of 
the observer. A significant feature of this kernel is that once the 
acceleration is turned
off at $\tau =\tau_1$, then for $\tau >\tau_1$,
\begin{equation}\label{eq8} F(\tau )=\hat{f}(\tau 
)+\int^{\tau_1}_{\tau _0}k_0(\tau ')\hat{f}(\tau ')\,d\tau'.
\end{equation}
There is therefore a {\it constant} memory of past acceleration and 
the field $F$ satisfies the standard field equations in the inertial 
frame. That is, the field
equations are linear differential equations and the addition of a 
constant solution is always permissible but subject to boundary 
conditions. In terms of actual
laboratory devices that have experienced accelerations in the past, 
such constant fields as in equation~\eqref{eq8} would be canceled 
once the devices are reset.
Thus case~(1) involves simple ``nonpersistent'' memory of past 
acceleration; therefore, we call $k_0$ the {\it kinetic memory} 
kernel. 

It is interesting to note that our basic integral 
equation~\eqref{eq2} together with the kinetic memory kernel~\eqref{eq7}
and an integration by parts takes the form 
\[
F(\tau)=F(\tau_0)+\int_{[\tau_0,\tau]} \Lambda df,
\]
so that $dF=\Lambda df$ along the worldline of the accelerated observer.
\subsection{Dynamic memory}
The second case involves a convolution type kernel 
$K=k(\tau -\tau ')$. It follows 
(see Appendix A) that in this case the resolvent 
kernel is of convolution type
as well, $R=r(\tau -\tau')$. Thus equation~\eqref{eq5} can be 
written, after expressing the left side as the area under the 
graph of the function $r$ from the origin to
$\tau-\tau_0=t$, as
\begin{equation}
\label{eq9}r(t)=\frac{d\Lambda 
(t+\tau_0)}{dt}\Lambda_0^{-1}.
\end{equation}
The kernel $k$ is then given by (cf. Appendix A)
\begin{equation}\label{eq10} k(t)=-r (t)+r\ast r(t)-r\ast r\ast 
r(t)+\cdots ,\end{equation}
where a star denotes the convolution operation. We note that in this 
case the {\it resolvent} kernel is directly proportional to 
acceleration, so that $r=0$ and,
by equation~\eqref{eq10}, $k=0$ for $t<0$ or $\tau <\tau_0$, i.e. 
before the acceleration is turned on. However, the character of 
memory that is indicated by $k$,
\begin{equation}\begin{split}\label{eq11} F(\tau )&=\hat{f}(\tau 
)+\int^\tau_{\tau_0}k(\tau -\tau ')\hat{f}(\tau ')\,d\tau '\\
&=\hat{f}(\tau)+\int^{\tau-\tau_0}_0k(t)\hat{f}(\tau 
-t)\,dt,\end{split}\end{equation}
is more complicated than in case~(1) 
due to the intricate relationship between $r(t)$ 
and $k(t)$ in equation~\eqref{eq10}. Even if the acceleration is 
turned off at $\tau =\tau_1$,
it turns out that $k$ does not vanish in general for $\tau>\tau_1$
and could even be divergent; in fact, proving the latter point is the 
main purpose of this paper.

Imagine, for instance, that $k(t)$ is finite everywhere and decays 
exponentially to zero for $t\to \infty$. Then in 
equation~\eqref{eq11}, as $\tau \to \infty$
long after the acceleration has been turned off at $\tau =\tau_1$, 
the contribution of the nonlocal term in \eqref{eq11} rapidly 
approaches a constant and we essentially
recover the ``nonpersistent'' kinetic memory familiar from 
case~(1). It turns out, however, that in general 
case~(2) involves situations with persistent
or {\it dynamic} memory such that under certain conditions $k(t)$ 
could diverge resulting in an asymptotically divergent $F(\tau )$.

The convolution (Faltung) type kernel is generally employed in many 
branches of physics and mathematics. As in equation~\eqref{eq11}, to 
produce the nonlocal part
of the output $F(\tau )$, an input signal $\hat{f}(\tau-t)$ is 
linearly folded, starting from $\tau$ and going backwards in proper 
time until $\tau_0$, with a
weight function $k(t)$ that is the impulse response of the system. 
The use of convolution type kernels is standard practice in 
phenomenological treatments of the
electrodynamics of media~\cite{7,7a,7b}, feedback control systems~\cite{8}, 
etc. We find, however, that for the pure vacuum case the convolution 
kernel due to \emph{nonuniform acceleration} 
in general leads to instability and is 
therefore unacceptable. This proposition is proved in the following 
section for the translational and rotational 
accelerations of the observer. 

The simplicity of the kinetic memory versus dynamic memory has been
particularly stressed by Hehl and Obukhov in their investigations of
nonlocal electrodynamics~\cite{ho,hoa}; moreover, their work has led
to the question of the ultimate physical significance of the convolution type
kernel in the nonlocal theory of accelerated systems~\cite{ho,hoa}. 
This question is settled in the present paper in favor of the
kinetic memory kernel.
\section{Dynamic memory of accelerated motion}
\subsection{Linear acceleration}

Imagine an observer at rest on the $z$-axis for $-\infty <\tau 
<\tau_0$. At $\tau =\tau _0$, the observer accelerates along the 
positive $z$-direction with
acceleration $g(\tau )>0$. For $\tau\ge \tau_0$, we set
\begin{equation}\label{eq12}
\theta (\tau )=\int^\tau_{\tau_0}g(\tau ')\,d\tau ',\end{equation}
$C=\cosh \theta$ and $S=\sinh \theta$. The natural nonrotating 
orthonormal tetrad frame of the observer along its worldline is given 
by
\begin{equation}\begin{array}{ll} \hat{\lambda}^\mu_{(0)} = (C,0,0,S), & 
\hat{\lambda}^\mu_{(1)}=(0,1,0,0),\\
\hat{\lambda}^\mu_{(2)}=(0,0,1,0), & 
\hat{\lambda}^\mu_{(3)}=(S,0,0,C).\end{array}\end{equation}
In this case $\Lambda (\tau )$ is given by
\begin{equation}\label{eq14} \Lambda =\begin{bmatrix} U & V\\-V 
&U\end{bmatrix},\quad U=\begin{bmatrix} C & 0 & 0\\ 0 & C & 0\\0 & 0 
& 1\end{bmatrix},\quad
V=SI_3,\end{equation}
where $I_i$, $(I_i)_{jk}=-\epsilon_{ijk}$, is a $3\times 3$ matrix 
proportional to the operator of infinitesimal rotations about the 
$x^i$-axis. 

Let us first consider case~(1), for which the kernel can be 
easily computed using equation~\eqref{eq7},
\begin{equation}\label{eq15} k_0(\tau )=-g(\tau )\begin{bmatrix}0 & 
I_3\\ -I_3 & 0\end{bmatrix},\end{equation}
so that when the acceleration is turned off at $\tau =\tau_1$ the 
kernel $k_0$ vanishes with the acceleration for $\tau \geq \tau_1$. 
On the other hand, $k_0$ is
simply constant for uniform acceleration (i.e. hyperbolic motion) 
with $g(\tau )=g_0$ for $\tau \geq \tau_0$. In the rest of this 
section, we focus attention on
case~(2) involving the convolution kernel.

For the convolution kernel, the resolvent kernel is given, via 
equation~\eqref{eq9}, by
\begin{equation}\label{eq16} r(\tau -\tau_0)=g(\tau )\begin{bmatrix} 
SJ_3 & CI_3\\-CI_3 & SJ_3\end{bmatrix},\end{equation}
where $(J_k)_{ij}=\delta_{ij}-\delta_{ik}\delta_{jk}$. In principle, 
the convolution kernel can be computed via the substitution of 
equation~\eqref{eq16} in
equation~\eqref{eq10}; however, this turns out to be  a daunting task in 
practice. Imagine, for instance, that the acceleration is turned off 
at $\tau=\tau_1$, so
that the resolvent kernel \eqref{eq16} has compact support over a 
time interval of length $\alpha=\tau_1-\tau_0$ and vanishes 
otherwise. It then follows that the
$r^{\ast\, n}$ term in the expansion \eqref{eq10} has compact support over 
a time interval of length $n\alpha$. The summation of series 
\eqref{eq10} turns out to be
rather complicated, except for the case of {\it uniform} 
acceleration, i.e. $g(\tau )=g_0$ for $\tau \geq \tau_0$, and the 
result is
\begin{equation}\label{eq17} k=-g_0\begin{bmatrix} 0 & I_3\\ -I_3 & 
0\end{bmatrix}.\end{equation}
It is interesting to note that equation \eqref{eq17} is the \emph{same} as 
the result of case~(1), equation~\eqref{eq15}, for \emph{uniform}
acceleration.

In view of the difficulty of summing the series \eqref{eq10} 
directly, we find it advantageous to use Laplace transforms, which we 
denote by an overbar, i.e.
$\mathcal{L} \{ k(t)\}=\bar{k}(s)$, where
\begin{equation}\label{eq18} \bar{k} (s)=\int^\infty_0 
e^{-st}k(t)\,dt;\end{equation}
then, taking the Laplace transform of equation~\eqref{eq10} and using 
the convolution (Faltung) theorem repeatedly, we arrive at
\begin{equation}\label{eq19} \bar{k} (s)=[I+\bar{r} (s)]^{-1}-I,\end{equation}
which is consistent with the reciprocity between $k$ and $r$.

\subsection{Stepwise acceleration}\label{sec:sa}
\begin{figure}[tb]  
\centerline{\psfig{file=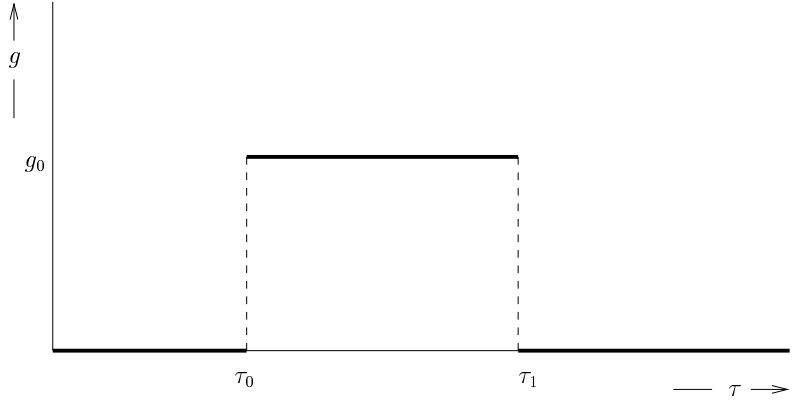, width=25pc}}
\caption{\label{fig1}
The linear acceleration of an observer that undergoes uniform
acceleration $g_0$ during a period $\alpha=\tau_1-\tau_2$ of its
proper time. If the area under the graph exceeds a critical value given
by $\beta_0\approx 1.2931$,
then the convolution kernel leads to divergences.}
\end{figure}
Let us specialize to a simple case of stepwise 
uniform acceleration, namely, we let $g(\tau)=g_0$ for $\tau_0\leq 
\tau \leq \tau_1$ and zero
otherwise (see Figure~\ref{fig1}). In this case,
\begin{equation} \label{eq20} \bar{r} (s)=\begin{bmatrix} \bar{r}_1 & 
\bar{r}_2\\ -\bar{r}_2 & \bar{r}_1\end{bmatrix},\end{equation}
where $\bar{r}_1(s)=q(s)J_3$ and $\bar{r}_2(s)=p(s)I_3$. Here 
$p(s)=\mathcal{L}\{gC\}$ and $q(s)=\mathcal{L}\{ gS\}$. All $6\times 
6$ matrices that we consider in
this paper have the general form \eqref{eq20}, i.e. each is 
completely determined by two $3\times 3$ matrices just as $\bar{r}_1$ 
and $\bar{r}_2$ characterize
$\bar{r}$ in equation~\eqref{eq20}; we therefore write $\bar{r}\to 
[\bar{r}_1;\bar{r}_2]$ to express this decomposition as
in equation~\eqref{eq20}. To find the Laplace 
transforms of $gC$ and $gS$, we
note that in equation~\eqref{eq12}, $\theta=g_0(\tau-\tau_0)$ for 
$\tau \leq \tau_1$ and $\theta=\beta_0=g_0(\tau _1-\tau_0)$ for
$\tau \geq \tau_1$; therefore,
\begin{equation}\label{eq21} p(s)\pm q(s)=\frac{g_0}{s\mp 
g_0}[1-e^{-(s\mp g_0)\alpha}],\end{equation}
where $\alpha = \tau_1-\tau_0=\beta_0/g_0$ is the acceleration time 
interval. Using the results of Appendix B, we find from
equation~\eqref{eq19} that $\bar{k}(s)$ 
can be expressed as
\[\bar{k}(s)\to [\beta_0Q(s)J_3;\beta_0P(s)I_3],\] where
\begin{align}\label{eq22} 
P(s)&=\frac{e^w}{D}[w(-e^w+\cosh \beta_0)+\beta_0\sinh \beta_0],\\
\label{eq23} Q(s) &= \frac{1}{D} [e^w(w\sinh \beta_0-\beta_0\cosh 
\beta_0)+\beta_0].
\end{align}
Here $w:=s\alpha$ and the denominator $D$ can be factorized as
\begin{equation}\label{eq24} 
D=(we^w-\beta_0e^{\beta_0})(we^w+\beta_0e^{-\beta_0}).
\end{equation}

It is useful to recall that the kernel $k\to [k_1;k_2]$ 
refers to a system at rest 
on the $z$-axis for $\tau \leq \tau_0$ that is uniformly accelerated
at $\tau=\tau_0$ 
with acceleration $g_0$
until $\tau_0+\alpha =\tau_1$, and then continues with uniform speed 
$\tanh \beta_0$ along the positive $z$-direction for $\tau \geq 
\tau_1$. Under certain
conditions, it is possible to obtain series representations for $k_1$ 
and $k_2$ (see Appendix C); however, to gain insight into the 
asymptotic behavior of $k_1$
and $k_2$ it proves more fruitful to proceed with an investigation of 
the singularities of $\bar{k}_1(s)=\beta_0 Q(s) J_3$ and 
$\bar{k}_2(s)=\beta_0 P(s)I_3$ in the complex 
$s$-plane. 
This is due to a simple property of the Laplace transformation in 
equation~\eqref{eq18} extended to the complex $s$-plane: let us
suppose that the convolution kernel $k(t)$ is a bounded function
for all $t=\tau-\tau_0>0$ as one naturally expects of a function
that represents memory; then, for any $s$ in the complex plane with
positive real part, i.e. $\re(s)>0$, equation~\eqref{eq18} implies
that the absolute magnitude of $\bar k(s)$ should be finite,
i.e. $\bar k(s)$ cannot be singular. Therefore, if we could show that
$\bar k(s)$ has in fact pole singularities at complex values of $s$ 
with $\re(s)>0$, then it would simply follow that 
$k(t)$ cannot be bounded for all $t>0$ and would thus be unsuitable
to represent the memory of finite accelerated motion. 

We will prove the following result: If $\beta_0 \exp(\beta_0)> 3\pi/2$,
then the corresponding function $k$ is unbounded for $t\ge 0$.
It suffices to show that $\bar k$ has a pole in the right half of
the complex $s$-plane. In fact, let us suppose that $\bar k$ has a pole
at $s=s_0$, where $\re (s_0)>0$,
but  $\norm{k}:=\sup_{t\ge 0} \abs{k(t)}<\infty$.  
In this case, $\bar k$ has a pole in the half-plane
$\calH$ consisting of all complex numbers $s$ such that 
$\re(s)\ge \frac{1}{2} \re(s_0)$, and therefore $\abs {\bar k}$ is not
bounded on $\calH$. On the other hand, for $s\in \calH$,
we have that
\[
\abs{\bar k(s)}\le \int_0^\infty e^{-\re(s) t}\abs{k(t)}\,dt
\le \norm{k}\int_0^\infty e^{-\re(s_0) t/2}\,dt<\infty,
\]
in contradiction. Thus the rest of this subsection is devoted
to the
determination of the poles of $\bar k(s)$ in the right half-plane.

The poles of $\bar k$ are elements of the zero set of $D$ with
$w=s \alpha$ and $\alpha>0$. Note, however,
that the (real) zeros $w=\pm\beta_0$ are removable singularities. 
Poles in the right half-plane are the zeros of $D$ with nonzero
imaginary parts. In view of the definition of $D$, let us consider
the complex roots in the right half-plane of the 
equation $w\exp(w)=b$, 
where $b$ is one of the real numbers $\pm\beta_0\exp(\pm \beta_0)$.
Because the zero set of this relation is symmetric with respect
to the real axis, it suffices to consider only roots in the
first quadrant of the complex $w$-plane.

We set $w=\xi+i\eta$, where $\xi\ge 0$  and $\eta\ge 0$ are real variables,
and note that $w\exp(w)=b$ if and only if
\[
\xi e^\xi=b\cos\eta,\qquad \eta e^\xi=-b \sin\eta.
\] 
If this system of equations has a solution,
then,  by squaring, adding and rearranging,
we have that  
$
\eta^2=b^2 \exp(-2\xi)-\xi^2
$
or, since $\eta\ge 0$, $\eta=\sqrt{b^2 \exp(-2\xi)-\xi^2}$.

There are several cases. For example, 
for $b>0$, there is a pole in the right half-plane
if the system of equations
\[
\eta=\sqrt{b^2 e^{-2\xi}-\xi^2},\qquad \xi e^\xi=b\cos\eta
\]
has a solution with $\xi>0$ and $\eta\bmod 2\pi\in (3\pi/2,2\pi)$.
Similarly, for $b<0$, there is a pole in the right half-plane
if the system of equations
\[
\eta=\sqrt{b^2 e^{-2\xi}-\xi^2},\qquad \xi e^\xi=b\cos\eta
\]
has a solution with $\xi>0$ and $\eta\bmod 2\pi\in (\pi/2,\pi)$.

A necessary condition for  the relation
$\eta=\sqrt{b^2 \exp(-2\xi)-\xi^2}$ to have a solution $(\xi,\eta)$
is  that $\xi \exp(\xi)<\abs{b}$. 
For  $b<0$, we must have $\xi \exp(\xi)<\beta_0\exp(-\beta_0)$; hence,
there is a unique real number $\xi_0$ such that the necessary condition
is met whenever $\xi\le \xi_0$. 
On the other hand, for $b>0$, the necessary condition,
$\xi \exp(\xi)<\beta_0\exp(\beta_0)$, is met if and only if $\xi<\xi_0=\beta_0$. 

Let us view $\eta$ as a function of $\xi$ and note that 
$\eta(0)=\abs{b}$, $\eta(\xi_0)=0$, and 
\[
\eta\frac{d\eta}{d\xi}=-e^{-2\xi}b^2-\xi<0
\]
for $\xi\ge 0$. In particular, $\eta$ decreases monotonically for
$0\le \xi\le \xi_0$.

Consider the relation $\xi \exp(\xi)=b\cos\eta$.
At $\xi=0$,  we have $\cos\eta=0$; therefore, the implicitly
defined function $\eta$ is such that 
$\eta(0)$ is an odd integer multiple of $\pi/2$. 
At $\xi_0$, we have $\cos\eta=\pm 1$ according to the sign
of $b$. In fact, $\eta(\xi_0)$ is an even multiple of
$\pi$ for $b>0$ and an odd multiple of $\pi$ for $b<0$.
Also, let us note that
\[
(\xi+1)e^\xi=-b\frac{d\eta}{d\xi}\sin\eta .
\]

\begin{figure}[tb]  
\centerline{\psfig{file=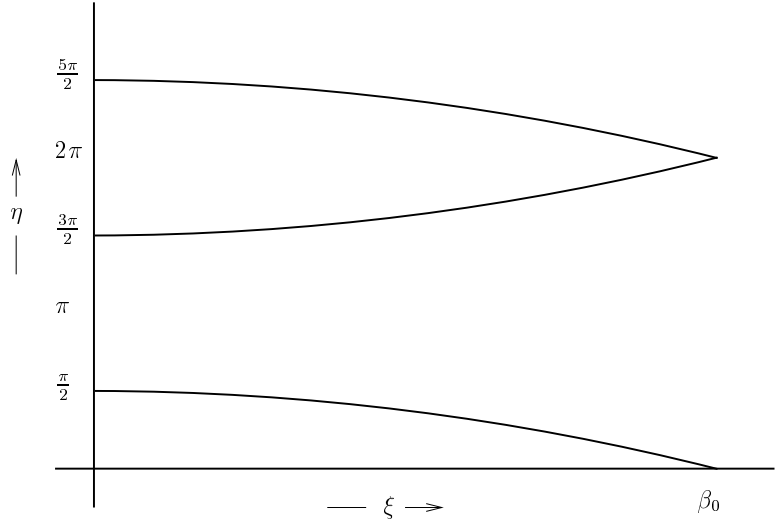, width=25pc}}
\caption{\label{fig2}The real branches of $\xi \exp(\xi)=b\cos\eta$
for $b>0$.}
\end{figure}
Suppose that $b>0$. We will determine the positions of the
real branches of the curve defined
by $\xi \exp(\xi)=b\cos\eta$. For $0\le \eta\le \pi/2$, we have
$\sin\eta>0$ and $d\eta/d\xi<0$, so there is a real branch
connecting the points $(0,\pi/2)$ and $(\xi_0,0)$ in the 
$(\xi,\eta)$-plane. For $\pi/2<\eta< 3\pi/2$, we have
$\cos\eta<0$; thus, there is no real branch in this region.
There is a real branch connecting $(0,3\pi/2)$ and $(\xi_0,2\pi)$
with $d\eta/d\xi>0$. 
This pattern continues as depicted in Figure~\ref{fig2}.
Note, however, that only the ``increasing'' branches correspond to
poles in the right half-plane. Indeed, for $b>0$, it is necessary that 
$\eta\bmod 2\pi$ be in the interval $(3\pi/2,2\pi)$. In particular,
the ``lowest'' branch corresponding to a pole connects
the points $(0,3\pi/2)$ and $(\xi_0,2\pi)$. It is now clear that
the curve defined by $\eta=\sqrt{b^2 \exp(-2\xi)-\xi^2}$ intersects
an increasing branch with $\xi>0$ if and only if $b>3\pi/2$.
The number of poles in the right half-plane increases by one  
as $b$ increases past an odd multiple of $\pi/2$.

\begin{figure}[tb]  
\centerline{\psfig{file=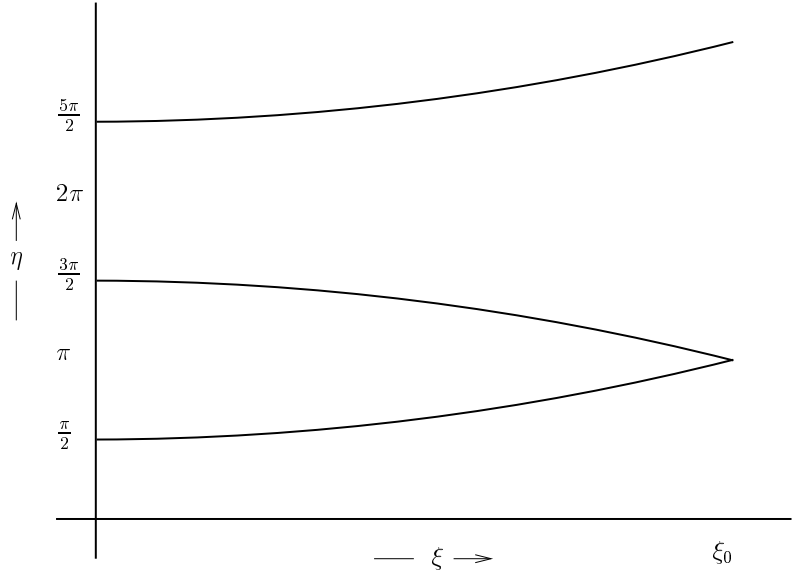, width=25pc}}
\caption{\label{fig3}The real branches of $\xi \exp(\xi)=b\cos\eta$
for $b<0$.}
\end{figure}
Suppose that $b<0$. In this case, the real branches of 
$\xi \exp(\xi)=b\cos\eta$
exist only if $\cos\eta<0$ as in Figure~\ref{fig3} and  
a corresponding
pole in the open right half-plane does not exist unless $\abs{b}>\pi/2$. 
Using the definition of $b$, 
this condition is equivalent to the requirement
that $\beta_0\exp(-\beta_0)>\pi/2$. But, the maximum value
of  $\beta_0\exp(-\beta_0)$ is $1/e<\pi/2$. Hence, negative values of
$b$ do not correspond to poles in the right half-plane.

We conclude that the dynamic memory kernel $k$ for
stepwise uniform linear acceleration is unbounded for 
$\beta_0=g_0\alpha>1.3$.

\subsection{Rotation}
\begin{figure}[tb]  
\centerline{\psfig{file=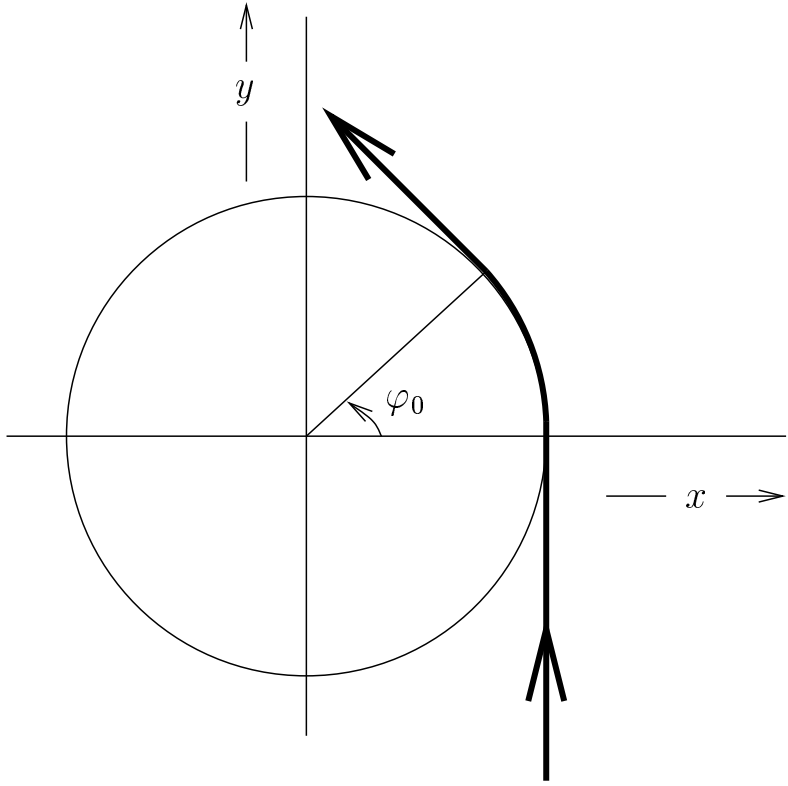, width=25pc}}
\caption{\label{fig4} Schematic plot of the motion of the observer
that undergoes stepwise uniform rotation of frequency $\Omega_0$ during
a period $\alpha=\tau_1-\tau_2$ of its proper time such that 
$\vp=\gamma \Omega_0\alpha$. If $\vp$ exceeds $\pi/2$,
then the convolution kernel leads to divergences.}
\end{figure}
Imagine next an observer that is initially moving uniformly with speed $v$ 
in the $(x,y)$-plane along a line parallel to the $y$-axis at 
$x=\rho_0$. At $t=0$,
$x=\rho_0$ and $y=0$, the observer starts rotating on a circle of 
radius $\rho_0$ with uniform frequency $\Omega_0=v/\rho_0$ in the 
positive sense around the
$z$-axis. Though the motion is continuous, there is no acceleration 
for $t<0$ and uniform circular acceleration for $t>0$. The natural 
orthonormal tetrad frame of
the uniformly rotating observer is given by
\begin{equation}\begin{split} \hat{\lambda}^\mu_{(0)}&=\gamma 
(1,-v\sin \varphi ,v\cos \varphi ,0),\\
\hat{\lambda}^\mu _{(1)}&= (0,\cos \varphi ,\sin \varphi ,0),\\
\hat{\lambda}^\mu_{(2)}&= \gamma (v,-\sin \varphi ,\cos \varphi ,0),\\
\hat{\lambda}^\mu _{(3)}&= (0,0,0,1),\label{eq25}\end{split}\end{equation}
where $\gamma =(1-v^2)^{-\frac{1}{2}}$ is the Lorentz factor and 
$\varphi =\Omega_0t=\gamma \Omega_0\tau$, so that we have set 
$\tau_0=0$ in this case. Computing
$\phi_{\alpha \beta}$ for the tetrad frame \eqref{eq25}, we find as 
expected that the translational acceleration has only a radial 
component
$g_1=-v\gamma^2\Omega_0$ and the rotational frequency is along the 
$z$-direction with magnitude $\Omega_3=\gamma^2\Omega_0$. Thus 
$\hat{f}=\Lambda f$, where
$\Lambda \to [\Lambda_1;\Lambda_2]$ is given by
\begin{equation}\label{eq26} \Lambda_1 =\begin{bmatrix} \gamma \cos 
\varphi & \gamma \sin \varphi & 0\\ -\sin \varphi & \cos \varphi &0\\ 
0 & 0 &\gamma
\end{bmatrix},\quad \Lambda_2=v\gamma \begin{bmatrix} 0 & 0 &1\\ 0 & 
0 & 0\\ -\cos \varphi & -\sin \varphi & 0\end{bmatrix} .\end{equation}

Let us first consider case~(1); the kinetic memory kernel 
$k_0$ can be easily computed using the fact that for $\Lambda$
given by equation~\eqref{eq26} we have $\Lambda^{-1}\to 
[\Lambda^T_1;\Lambda_2^T]$. Then
we find that $k_0\to 
[$\boldmath$\Omega$\unboldmath$\cdot$\boldmath$I$\unboldmath$;-\mathbf{g}\cdot 
\mathbf{I}]$, where
\boldmath$\Omega$\unboldmath$=(0,0,\gamma^2\Omega_0)$ and $\mathbf{g} 
=(-v\gamma^2\Omega_0,0,0)$ with respect to the orthonormal tetrad 
frame \eqref{eq25}. Thus
$k_0$ is a constant kernel so long as the observer rotates uniformly; 
for instance, if the acceleration is turned off at $\tau_1=\alpha$ 
corresponding to
$\vp=\gamma \Omega _0\alpha$, then the observer will have 
uniform linear motion again with speed $v$ for $\tau >\tau_1$ and the 
kernel $k_0$ will vanish (see Figure~\ref{fig4}).

Let us now consider case~(2); the dynamic memory kernel is 
given by the series \eqref{eq10} in terms of the resolvent kernel. 
This is given by
equation~\eqref{eq9}, $r\to [r_1;r_2]$, where
\begin{equation}\begin{split}\label{eq27} r_1&=\gamma 
\Omega_0\begin{bmatrix} -\gamma^2 \sin \varphi & \gamma \cos \varphi 
&0\\ -\gamma \cos \varphi & -\sin \varphi
& 0\\ 0& 0 & v^2\gamma^2\sin \varphi \end{bmatrix},\\ r_2&= 
v\gamma^2\Omega_0\begin{bmatrix} 0 & 0 & \gamma \sin \varphi\\ 0 & 0 
& \cos \varphi \\ \gamma \sin
\varphi & -\cos
\varphi & 0\end{bmatrix}.\end{split}\end{equation}
The explicit calculation of $k$ using the series \eqref{eq10} for the 
general case of stepwise uniform rotation from $\tau =0$ to 
$\tau_1=\alpha$ is rather
complicated; however, for $\tau_1\to \infty$ the calculation can be 
carried through and the result is a constant kernel given by $k\to
[$\boldmath$\Omega$\unboldmath$\cdot \mathbf{I};-\mathbf{g}\cdot 
\mathbf{I}]$. Just as in the case of uniform translational 
acceleration (cf. Section 4), we have
$k_0=k$ for uniform rotation as well.

To calculate $k$ for the stepwise uniform rotation of duration 
$\tau_1-\tau_0=\alpha >0$, we use Laplace transforms as in the 
previous section (see Figure~\ref{fig4}). Let
$C'=\alpha^{-1}\mathcal{L}\{\cos \varphi\}$ and $S'=\alpha 
^{-1}\mathcal{L} \{ \sin \varphi \}$; then, with $w=s\alpha$ we find
\begin{equation}\label{eq28}
 C'\pm i S'= 
\frac{1-e^{-(w\mp i\vp)}}{w\mp i\vp},
\end{equation}
and hence the Laplace transform of the resolvent kernel is given by 
$\bar{r}\to [\bar{r}_1;\bar{r}_2]$, where
\begin{equation}\label{eq30} \bar{r}_1=\vp\begin{bmatrix} 
-\gamma^2S' & \gamma C' & 0\\ -\gamma C' & -S' & 0\\ 0 & 0 & 
v^2\gamma^2S'\end{bmatrix},\quad
\bar{r}_2=v\gamma \vp\begin{bmatrix} 0 & 0 & \gamma S'\\ 0 & 0 
& C'\\ \gamma S' & -C' & 0\end{bmatrix}.\end{equation}
Using methods given in Appendix B, equation~\eqref{eq19} leads to 
\[\bar{k}(s)\to [\bar{k}_1(s);\bar{k}_2(s)],\]
where
\begin{eqnarray}
\label{eq31} 
 \bar{k}_1(s)&=&\vp
\begin{bmatrix} 
\gamma^2 \mathcal{Q} & \gamma \mathcal{P} & 0\\ 
-\gamma \mathcal{P} & \mathcal{Q} & 0\\
 0 & 0 & -v^2\gamma^2\mathcal{Q}
\end{bmatrix},\\ 
\label{eq31b}\bar{k}_2(s)&=&v\gamma \vp
\begin{bmatrix} 
0 & 0 & -\gamma \mathcal{Q}\\
 0 & 0 & \mathcal{P}\\
-\gamma \mathcal{Q} & -\mathcal{P} & 0
\end{bmatrix}.
\end{eqnarray}
Here $\mathcal{P}$ and $\mathcal{Q}$ are given by
\begin{align}\label{eq32} \mathcal{P} &= 
\frac{e^w}{\mathcal{D}}[w(-e^w+\cos \vp)-\vp\sin 
\vp],\\
\label{eq33} \mathcal{Q} &= \frac{1}{\mathcal{D}} [e^w(-w\sin 
\vp+\vp\cos \vp)-\vp],\end{align}
and the denominator $\mathcal{D}$ is given by
\begin{equation}\label{eq34} 
\mathcal{D}=(we^w-i\vp e^{i\vp})
                   (we^w+i\vp e^{-i \vp}).
\end{equation}
It is interesting to note that if we formally substitute $\beta_0$ 
for $i\vp$ in equations~\eqref{eq32}--\eqref{eq34}, we obtain 
results familiar from the previous subsection;
specifically, under $i\vp\to \beta_0$, $\mathcal{P}\to P$, 
$\mathcal{Q}\to iQ$ and $\mathcal{D}\to D$, where $P,Q$ and $D$ are 
given in
equations~\eqref{eq22}--\eqref{eq24}. Therefore, the main results of 
the previous subsection
can also be used in the analysis of stepwise uniform rotation; 
for instance, with appropriate 
modifications the explicit
expressions given in Appendix C for the convolution kernel in a 
special case can be employed here as well. However, since 
$\vp>0$, the singularities of
$\mathcal{P}$ and $\mathcal{Q}$ are in general different from those 
in the previous subsection.

To determine the pole singularities of $\bar k(s)$ in
the right half-plane in the case of stepwise rotation 
it suffices to consider the equation 
\begin{equation}\label{eq:our}
we^w= i\vp e^{i \vp}
\end{equation}
with $\vp>0$.
Indeed, note that if $w$ is a solution of this equation,
then the complex conjugate of $w$ is a solution of 
$w\exp(w)= -i\vp \exp(-i \vp)$.

As before, let us set 
$w=\xi+i\eta$ and note that equation~\eqref{eq:our} 
is equivalent to the system
of real equations given by
\begin{equation}\label{eq:psign}
\xi e^\xi=\vp\sin(\eta-\vp),\qquad \eta e^\xi=\vp\cos(\eta-\vp),
\end{equation}
where $\xi\ge 0$ and $\vp>0$.
We recall here that the solution $w=i\vp$, i.e. $\xi=0$
and $\eta=\vp$, of equations~\eqref{eq:our} and~\eqref{eq:psign} 
corresponds to a removable singularity.
A necessary condition for system~\eqref{eq:psign}
to have a solution with $\xi>0$ is that $\sin(\eta-\vp)>0$ and
$\eta\cos(\eta-\vp)>0$; the latter condition means that
 $\cos(\eta-\vp)$ and $\eta$ must have the
same sign. 

Consider the system
\begin{equation}\label{eq:nsys}
(\xi^2+\eta^2) e^{2 \xi}=\vp^2,\qquad \xi e^\xi=\vp\sin(\eta-\vp).
\end{equation}
If it has a solution $(\xi,\eta)$, 
then it follows from display~\eqref{eq:nsys} that
\[
\eta^2 e^{2\xi}=\vp^2\cos^2(\eta-\vp),
\]
and therefore $\eta \exp(\xi)=\pm \vp\cos(\eta-\vp)$.
Comparing this result with system~\eqref{eq:psign}, we conclude
that we can use system~\eqref{eq:nsys} for finding the poles if we keep
in mind that  $\eta$ and
$\cos(\eta-\vp)$ must have the same sign. 

\begin{figure}[tb]  
\centerline{\psfig{file=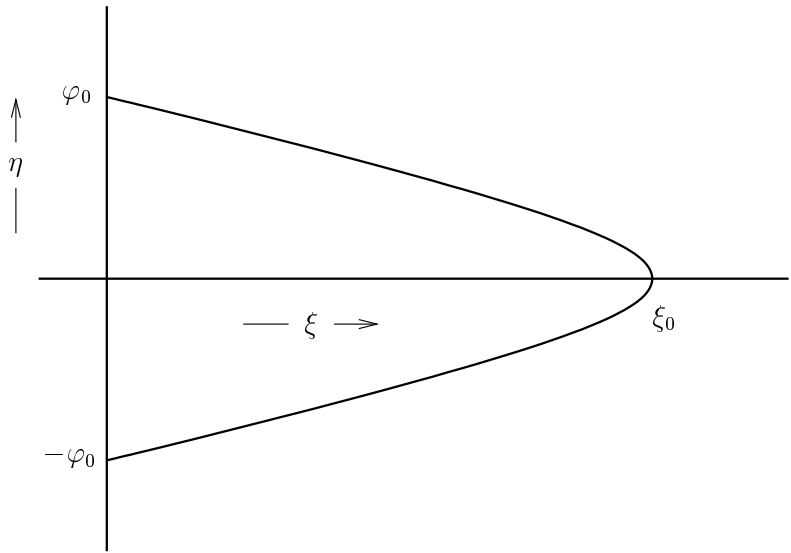, width=25pc}}
\caption{\label{fig5} The graph of $\eta^2=\vp^2\exp(-2\xi)-\xi^2$.}
\end{figure}
The first equation in display~\eqref{eq:nsys} is equivalent to
$\eta^2=\vp^2\exp(-2\xi)-\xi^2$. Its graph 
in the right half-plane has the form depicted in Figure~\ref{fig5},
where $\xi_0$ is the unique real solution of the equation 
$\xi \exp(\xi)=\vp$.

The poles we seek correspond to the intersections of the graph in 
Figure~\ref{fig5}
with the real branches of the second curve in display~\eqref{eq:nsys}.
The intercepts of these branches with the $\eta$-axis are given
by the solutions of the equation $\sin(\eta-\vp)=0$; that is,
$\eta$ is equal to $\vp$ plus an integer multiple of $\pi$. Along the
line given by $\xi=\xi_0$, the intercepts are given by 
$\xi_0 \exp(\xi_0)=\vp \sin(\eta-\vp)$. Because $\xi_0 \exp(\xi_0)=\vp$, 
these intercepts
are the solutions of $\sin(\eta-\vp)=1$; that is,
$\eta$ is $\vp+\pi/2$ plus an integer multiple of $2\pi$.
The shape of the branches connecting points on the two vertical lines
(at $\xi=0$ and $\xi=\xi_0$) is determined by the sign of $\cos(\eta-\vp)$ along the branch.
Indeed, we have already established that poles occur only
at points where $\eta$ and $\cos(\eta-\vp)$ have the same sign. 
Note that
\[
(\xi+1)e^\xi=\vp\frac{d\eta}{d\xi}\cos(\eta-\vp),
\]
and therefore the slope of the branch has the same sign as $\cos(\eta-\vp)$. 
Moreover, only the branches with $\sin(\eta-\vp)>0$ correspond
to poles in the right half-plane.

\begin{figure}[tb]  
\centerline{\psfig{file=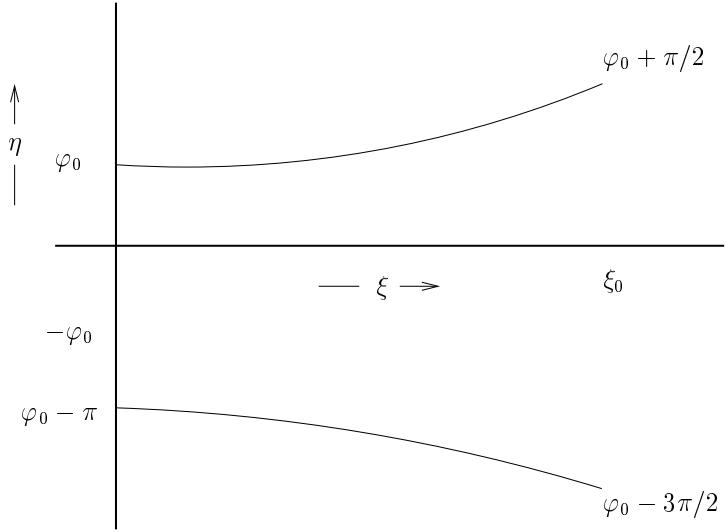, width=25pc}}
\caption{\label{fig6} The graph of $\xi \exp(\xi)=\vp \sin(\eta-\vp)$
for $0<\vp<\pi/2$.}
\end{figure}
There are several cases depending on the size of $\vp$.
For $0<\vp<\pi/2$, it is easy to see that the important branches
are as depicted in Figure~\ref{fig6}. These would not 
intersect the graph in Figure~\ref{fig5}; hence, there are no poles in
the right half-plane.

We will next show that if $\vp>\pi/2$, then there is at least one pole in
the right half-plane. For $\vp$ in this range, there is 
an integer  $j\ge 1$ such that $j\pi/2\le \vp< (j+1)\pi/2$. In particular,
we have that 
$\vp-j\pi/2\ge 0$ and $\vp-(j+1)\pi/2<0$. 
There are four cases.
(1) Suppose that $j$ is even and $\cos (j\pi/2)=1$. The branch of the
curve $\xi \exp(\xi)=\vp \sin(\eta-\vp)$ with $\eta$-intercept
$\vp-j\pi/2\ge 0$ has positive slope (like the upper
branch in Figure~\ref{fig6}). Because $\vp-j\pi/2<\vp$,
this branch intersects the curve depicted in Figure~\ref{fig5} 
in the upper half-plane. This point corresponds to a pole. Indeed,
at the point of intersection $\sin(\eta-\vp)>0$ and
$\eta\cos(\eta-\vp)>0$.
(2) Suppose that $j$ is even and $\cos (j\pi/2)=-1$. 
The branch with  $\eta$-intercept $\vp-j\pi/2\ge 0$ has negative
slope and meets the line $\xi=\xi_0$ with ordinate $\vp-(j+1)\pi/2<0$.
Hence, this branch intersects the curve depicted in 
Figure~\ref{fig5} 
in the lower half-plane. This point corresponds to a pole.
(3) Suppose that $j$ is odd and $\cos ((j+1)\pi/2)=1$. The
branch of the curve with $\eta$-intercept  $\vp-(j+1)\pi/2$ has
positive slope and it meets the curve depicted in Figure~\ref{fig5}
in the upper half-plane where the intersection point corresponds to a pole.
For the subcase where $j=3$ and $\vp=3\pi/2$, 
it is interesting to note that
$\eta=0$ and $\xi_0$, such that 
$\xi_0 \exp(\xi_0)=3\pi/2$, is the pole.
(4) Suppose that $j$ is odd and $\cos ((j+1)\pi/2)=-1$. 
The curve with $\eta$-intercept $\vp-(j+1)\pi/2$ has
negative slope and $-\vp\le \vp-(j+1)\pi/2$. Hence, this branch meets
the curve depicted in Figure~\ref{fig5} in the lower half-plane where
the intersection corresponds to a pole.

We conclude that the dynamic memory kernel $k$ for stepwise uniform
rotation is unbounded for $\vp=\gamma\Omega_0\alpha>\pi/2$.
\subsection{Smooth acceleration} 
We have demonstrated
that the convolution kernel $k$ is unbounded for certain
stepwise translational and rotational accelerations. 
Could this result be due to the discontinuities of these accelerations 
at $\tau_0$ and $\tau_1$? To prove that this is \emph{not} the case,
we are interested here instead in smooth accelerations that
closely approximate the stepwise ones already studied. The
translational and rotational cases are in fact closely related as
we have demonstrated; therefore
in this subsection we show the same result for the simpler case of
\emph{smooth} translational acceleration.

Let us consider an acceleration $g$ with compact
support in the interval $[\tau_0,\tau_1]$.
By the definition of $\Lambda$ and the choice of $g$,
the matrix $\Lambda(\tau_0)=\Lambda_0$ is the $6\times 6$ identity
matrix. Using this fact and  
equations~\eqref{eq9} and~\eqref{eq14}, we find that 
$ r(t)=[g(\tau)S(\tau) J_3;g(\tau)C(\tau) I_3]$, 
where $t=\tau-\tau_0$, $S(\tau)=\sinh\theta$, $C(\tau)=\cosh\theta$
and $\theta(\tau)$ is given by equation~\eqref{eq12}.  
It follows from equation~\eqref{eq18} that 
$ \bar r(s)=[\calS(s) J_3;\calC(s) I_3]$,
where
\[
\calC(s)\pm\calS(s)
 =\int_0^\infty e^{-st}g(t+\tau_0)e^{\pm\theta(t+\tau_0)}\,dt.
\]
Using equation~\eqref{eq19} and the results of Appendix B, we find that
the Laplace
transform of the convolution kernel $k$ is given by
$\bar k(s)=[\calH_1(s)J_3;\calH_2(s)I_3]$, where
\[
\calH_1(s)=\frac{1+\calS}{(1+\calS)^2-\calC^2}-1,\qquad
\calH_2(s)=- \frac{\calC}{(1+\calS)^2-\calC^2}.
\]

We are interested in the zeros of the denominator
\[(1+\calS)^2-\calC^2=(1+\calS+\calC)(1+\calS-\calC).\]
It suffices to demonstrate that $1+\calS+\calC$ has a zero in
the right half of the complex $s$-plane.
Because $g$ has compact support in the interval $[\tau_0,\tau_1]$,
the function $1+\calS+\calC$ is given by
\[
s\mapsto 1+\int_0^\alpha  
e^{-s t} e^{\int_0^t g(\sigma+\tau_0)\,d\sigma} g(t+\tau_0)\,dt,
\]
where $\alpha=\tau_1-\tau_0$.
If $g$ is the stepwise uniform acceleration
considered previously, then this function reduces to
\[ s\mapsto \frac{e^{-\alpha s}}{s-g_0}
( se^{\alpha s}- g_0 e^{\alpha g_0});
\] 
and, by the results in Subsection~\ref{sec:sa} for equation~\eqref{eq24},
if $\beta_0 \exp(\beta_0)>3\pi/2$, it
has a zero in the right half of the complex $s$-plane corresponding
to a pole of $\bar k$. 
We will show that such a pole persists for a smooth acceleration
that is sufficiently close to the stepwise acceleration.

For an arbitrary acceleration $g$ with support in
the interval $[\tau_0,\tau_1]$, we define
the associated real-valued function $\zeta$ on
the interval $[0,\alpha]$ given  by $\zeta(t)=g(t+\tau_0)$.
Also, recall that the  $L^1$-norm of a real-valued function $\upsilon$ 
defined on the interval
$[0,\alpha]$ is given by
\[\norm{\upsilon}_1:=\int_0^\alpha\abs{\upsilon(t)}\,dt.\] 

Suppose that $\zeta$ and $\upsilon$ are real-valued functions
defined on the interval $[0,\alpha]$ such that $\norm{\zeta}<\infty$ and
$\norm{\upsilon}_1<\infty$, and consider the  
complex-valued analytic functions $Z$ and $\Upsilon$ 
of the complex variable $s$ given by  
\begin{eqnarray*}
Z(s)&=& 
1+\int_0^\alpha  
e^{-s t} e^{\int_0^t \zeta(\sigma)\,d\sigma} \zeta(t)\,dt,\\
\Upsilon(s)&=& 
1+\int_0^\alpha  
e^{-s t} e^{\int_0^t \upsilon(\sigma)\,d\sigma} \upsilon(t)\,dt.
\end{eqnarray*}
We will prove the following proposition.
\emph{If $Z$ has a zero in the open right-half of the complex 
$s$-plane
and $\norm{\upsilon-\zeta}_1$ is sufficiently small,
then $\Upsilon$ has a zero in the open right-half of 
the complex $s$-plane.}
By a standard result from mathematical analysis (see~\cite{ll}), 
if $\zeta$
is an $L^1$ function
(for example, if $\zeta(t)=g(t+\tau_0)$ for
the stepwise acceleration $g$), then $\norm{\zeta-\upsilon}_1$
can be made as small as desired for
a $C^\infty$ function $\upsilon$. Hence, by the proposition,
there is a smooth acceleration with compact support such that 
its associated
convolution kernel is unbounded.

Our proof begins with two estimates.
For notational convenience, let us define
\[
\hat{\zeta}(t)=e^{\int_0^t \zeta(\sigma)\,d\sigma} \zeta(t),\qquad
\hat{\upsilon}(t)=e^{\int_0^t \upsilon(\sigma)\,d\sigma} \upsilon(t).
\]
The first estimate is 
\begin{equation}\label{est:1}
\abs{\Upsilon(s)-Z(s)}\le \norm{\hat\upsilon-\hat\zeta}_1
\end{equation}
for all $s$ such that $\re (s)\ge 0$. 
To prove it, note that
\[
\abs{\Upsilon(s)-Z(s)}\le 
\int_0^\alpha \abs{e^{-s t}}\abs{\hat\upsilon(t)-\hat\zeta(t)}\,dt.
\]
Because $\abs{\exp(-s t)}\le 1$ for $\re(s)\ge 0$, we have the
inequality
\[\abs{\Upsilon(s)-Z(s)}\le \norm{\hat\upsilon-\hat\zeta}_1\]
for all $s$ in the closed right half-plane. 
The second estimate is 
\begin{equation}\label{est:2}
\norm{\hat\upsilon-\hat\zeta}_1\le 
e^{\norm{\zeta}_1}(1+\alpha\norm{\zeta}) e^{\norm{\upsilon-\zeta}_1}
\norm{\upsilon-\zeta}_1.
\end{equation}
To prove it, we have the triangle law estimate
\begin{eqnarray}\label{est:tl}
\nonumber\abs{\hat\upsilon(t)-\hat\zeta(t)}&\le&
\abs {e^{\int_0^t \upsilon(\sigma)\,d\sigma}\upsilon(t)
-e^{\int_0^t \upsilon(\sigma)\,d\sigma}\zeta(t)}\\
\nonumber &&{}+\abs {e^{\int_0^t \upsilon(\sigma)\,d\sigma}\zeta(t)
-e^{\int_0^t \zeta(\sigma)\,d\sigma}\zeta(t)}\\
&\le & e^{\int_0^t \abs{\upsilon(\sigma)}\,d\sigma}\abs{\upsilon-\zeta}
+\abs{\zeta}
\abs{e^{\int_0^t \upsilon(\sigma)\,d\sigma}-e^{\int_0^t \zeta(\sigma)\,d\sigma}}
\end{eqnarray}
and, by the mean value theorem (applied to the exponential function),
the inequality
\[
\abs{e^{\int_0^t \upsilon(\sigma)\,d\sigma}-e^{\int_0^t \zeta(\sigma)\,d\sigma}}
\le e^\varsigma
\big|\int_0^t \upsilon(\sigma)\,d\sigma-\int_0^t \zeta(\sigma)\,d\sigma\big|,
\]
where $\varsigma$ is some number between 
$\int_0^t \upsilon(\sigma)\,d\sigma$ and $\int_0^t \zeta(\sigma)\,d\sigma$.
If $\varsigma\le 0$, then $\exp(\varsigma)<1$; and if
$\varsigma>0$, then $\varsigma<\max\{\norm{\upsilon}_1,\norm{\zeta}_1\}$.
Hence, 
\[e^\varsigma\le e^{\max\{\norm{\upsilon}_1,\norm{\zeta}_1\}}\]
and, because 
$\norm{\upsilon}_1\le \norm{\upsilon-\zeta}_1+ \norm{\zeta}_1$,
we have that
\[e^\varsigma\le e^{\norm{\zeta}_1}e^{\norm{\upsilon-\zeta}_1}.\]
Using this result and the estimate~\eqref{est:tl}, it follows that
\begin{eqnarray*}
\abs{\hat\upsilon(t)-\hat\zeta(t)}
&\le&
e^{\norm{\upsilon}_1}\abs{\upsilon(t)-\zeta(t)}
+\norm{\zeta}
e^{\norm{\zeta}_1}e^{\norm{\upsilon-\zeta}_1}
\int_0^\alpha\abs{\upsilon(\sigma)-\zeta(\sigma)}\,d\sigma\\
 &\le& e^{\norm{\zeta}_1}e^{\norm{\upsilon-\zeta}_1}
(\abs{\upsilon(t)-\zeta(t)}+\norm{\zeta}\norm{\upsilon-\zeta}_1).
\end{eqnarray*}
Therefore,
\begin{eqnarray*}
\norm{\hat\upsilon-\hat\zeta}_1
&=&\int_0^\alpha\abs{\hat\upsilon-\hat\zeta}\,dt\\
&\le& e^{\norm{\zeta}_1}e^{\norm{\upsilon-\zeta}_1}
(\norm{\upsilon-\zeta}_1+\alpha\norm{\zeta}\norm{\upsilon-\zeta}_1)
\end{eqnarray*}
and 
a rearrangement of the right-hand side of the last inequality
gives the desired result.

In the rest of this section, we let $\zeta$ represent the
stepwise uniform linear acceleration and $\upsilon$ the
smooth linear acceleration that approximates it sufficiently closely. 
We then
choose a circle centered at a zero of $Z$ in the open right half 
of the complex $s$-plane such that the circle does not pass through
a zero of $Z$ and such that the circle is contained in
the open right half-plane. Let $\kappa$, a complex-valued function 
defined on the interval $[0,2\pi]$, be a continuous
parametrization of this circle and define two new functions
$\kappa_Z$ and $\kappa_\Upsilon$ on this interval by
\[\kappa_Z(\vartheta)=Z(\kappa(\vartheta)),\qquad 
\kappa_\Upsilon(\vartheta)=\Upsilon(\kappa(\vartheta)).
\]
The images of these functions are closed curves in the complex
$s$-plane. 
In complex analysis, the principle of the argument theorem~\cite{hen}
for an analytic function $\Delta$ relates the winding number of the
image of $\kappa_\Delta$ with respect to the origin to
the number of zeros of the function $\Delta$ inside the circle,
provided that the circle does not pass through any zero of $\Delta$.
If we show that $\kappa_Z$ and $\kappa_\Upsilon$ are homotopic and therefore
have the same winding number with respect to the origin and that the circle
does not pass through a zero of $\Upsilon$, then $Z$ and $\Upsilon$ must
have the same number of zeros inside the circle.

We claim that if $\norm{\upsilon-\zeta}_1$ is 
sufficiently small, then the image of $\kappa$ does not pass through a zero
of $\Upsilon$. To prove the claim, note that 
\[ m:= \min\{\abs{\kappa_Z(\vartheta)}:0\le \vartheta\le 2\pi\}>0\]
(because $\kappa$ does not pass through a zero of $Z$) 
and using the triangle inequality
\[
0<m\le \abs{\kappa_\Upsilon(\vartheta)}+\norm{\kappa_\Upsilon-\kappa_Z},
\]
where $\norm{\kappa_\Upsilon-\kappa_Z}$ is the supremum
of $\abs{\kappa_\Upsilon(\vartheta)-\kappa_Z(\vartheta)}$ for
$0\le \vartheta\le 2 \pi$.
Using the estimates~\eqref{est:1} and~\eqref{est:2},
we have that
\begin{equation}\label{est:kk}
\abs{\kappa_\Upsilon(\vartheta)-\kappa_Z(\vartheta)}
\le e^{\norm{\zeta}_1}e^{\norm{\upsilon-\zeta}_1}
(1+\alpha\norm{\zeta})\norm{\upsilon-\zeta}_1.
\end{equation}
By the estimate~\eqref{est:kk}, $\norm{\kappa_\Upsilon-\kappa_Z}$
can be made small, say less than  
$m/2$,
by taking $\norm{\upsilon-\zeta}_1$ sufficiently small. 
For all 
$\upsilon$ satisfying this requirement, which we impose for
the remainder of the proof,  we have that 
$\abs{\kappa_\Upsilon(\vartheta)}>0$; that is, $\kappa$ does not pass through a zero
of $\Upsilon$.

It remains to show that 
$\kappa_\Upsilon$ is homotopic to $\kappa_Z$. Assuming this
homotopy relation, the  image
curves  of $\kappa_\Upsilon$ and $\kappa_Z$ would 
have the same winding number with respect to the origin.
By the choice of $\kappa$ and the argument principle (see~\cite{hen}), 
the curve
$\kappa_Z$ has a nonzero winding number. Hence, 
$\kappa_\Upsilon$ would have the same nonzero winding number. Again,
by the argument principle,  $\Upsilon$ must then have a zero in
the disk bounded by the circle parametrized by $\kappa$, which
is the desired result.

To complete the proof we need to show that  
$\kappa_\Upsilon$ and $\kappa_Z$
are indeed homotopic. Let $\bbC$ denote the complex numbers. 
We will show that 
$H:[0,1]\times[0,2\pi]\to \bbC\setminus\{0\}$  given by
\[
H(\sigma, \vartheta)=\kappa_Z(\vartheta)-
\sigma (\kappa_Z(\vartheta)-\kappa_\Upsilon(\vartheta))
\]
is the required homotopy. By inspection, $H$ is continuous,
$H(0,\vartheta)=\kappa_Z(\vartheta)$ and 
$H(1,\vartheta)=\kappa_\Upsilon(\vartheta)$. Hence, it suffices
to show that $H(\sigma,\vartheta)\ne 0$ for all 
$(\sigma,\vartheta)\in [0,1]\times[0,2\pi]$.
By our choice of $\upsilon$, we have that 
$\norm{\kappa_\Upsilon-\kappa_Z}<m/2$; therefore
\[
\abs{H(\sigma, \vartheta)}
\ge \abs{\kappa_Z(\vartheta)}-
\abs{\sigma}\abs{\kappa_\Upsilon(\vartheta)-\kappa_Z(\vartheta)}
\ge m-\norm{\kappa_\Upsilon-\kappa_Z}> m/2,
\]
as required.

We conclude that the dynamic memory kernel $k$ for the smooth
linear acceleration that closely approximates the stepwise
acceleration is unbounded if the area under the graph of $g(\tau)$
exceeds a critical value $\sim 1$.

\section{Discussion}

We have investigated the properties of the nonlocal kernel that is 
induced by accelerated motion in Minkowski spacetime. The physical 
principles outlined in this
paper do not completely determine the kernel; therefore, simplifying 
mathematical assumptions need to be introduced in order to identify a 
unique kernel. Two
possibilities have been explored in this work corresponding to 
kinetic memory $(k_0)$ and dynamic memory $(k)$. We show that for 
accelerated motion that is uniform
(linear or circular), the two kernels give the same constant result 
$k_0=k$. They differ, however, if the acceleration is turned off at a 
certain moment. We have
therefore studied piecewise uniform acceleration (linear and circular)
and have demonstrated that the dynamic memory (convolution) kernel 
could be divergent and is
therefore ruled out. Furthermore, 
this conclusion is shown to be 
independent of the stepwise character of the linear acceleration considered.

The use of convolution kernels is standard practice in the nonlocal
electrodynamics of continuous media, where it is assumed phenomenologically
that memory always fades. In our treatment of acceleration-induced
nonlocality in vacuum, however, the behavior of memory must be determined
from first principles. In this connection, the possible advantage of kinetic
memory in terms of its simplicity was first emphasized by Hehl and 
Obukhov~\cite{ho,hoa}.

The theory developed here is applicable to any basic field; however, 
for the sake of concreteness and in view of possible observational 
consequences, we employ
electromagnetic radiation fields throughout. A basic consequence of 
the nonlocal theory of accelerated systems is that it is incompatible 
with the existence of a
basic scalar field; that is, in this case $\Lambda (\tau )=1$, $k_0=0$ 
and the nonlocality disappears so that a basic scalar radiation field
can stay completely at rest with respect to a rotating observer 
in contradiction with our fundamental 
physical assumption. This prediction of the nonlocal theory is in 
agreement with present
experimental data. Further confrontation of the nonlocal theory with 
observation is urgently needed.

\appendix

\section*{Appendix A}

Consider an integral equation of the form
\begin{equation}\label{A1} \phi (x)=\psi (x)+\epsilon 
\int^x_a K(x,y)\phi (y)\,dy,\tag{A1}\end{equation}
where $\psi$ is a continuous function, the kernel
$K$ is continuous and $\epsilon$ is a constant parameter. 
There is a unique continuous resolvent
kernel $R$ such that
\begin{equation}\label{A2}\tag{A2} \psi (x)=\phi (x)+\epsilon 
\int^x_aR(x,y)\psi (y)\,dy.\end{equation}
In turn, $K$ can be 
thought of as the resolvent kernel for $R$; this follows from the 
complete reciprocity between $K$
and $R$.

The proof of the existence and uniqueness of the resolvent
kernel is by successive approximation. In fact, the solution $\phi$ 
can be obtained as the uniform limit of the sequence of 
continuous functions
$\{\phi_n\}_{n=0}^\infty$  defined as follows:
$\phi_0 (x)=\psi(x)$ and 
\begin{equation}\label{A3}\tag{A3} \phi_{n+1}(x)=\psi (x)+\epsilon 
\int^x_a K(x,y)\phi_n(y)\,dy.\end{equation}
Thus
\begin{align}\label{A4}\tag{A4} 
\phi_1(x)=&\; \psi (x)+\epsilon \int^x_aK(x,y)\psi (y)\, dy,\\
 \nonumber \phi_2 (x)=&\;\psi (x)+\epsilon 
\int^x_aK(x,y)[\psi (y)+\epsilon \int^y_a K(y,z)\psi (z)\,dz]\,dy\\
\label{A5}
\tag{A5}
=&\;\phi_1(x)+\epsilon^2\int^x_aK(x,y)\int^y_a K(y,z)\psi (z)\, dz dy.
\end{align}
The integration in \eqref{A5} is over a triangular domain in the 
$(y,z)$-plane defined by the vertices $(a,a)$, $(x,a)$ and $(x,x)$. 
Changing the order of the
integration in \eqref{A5} results in the equality
\begin{equation*}\label{A6}\tag{A6}\begin{split} 
&\int^x_aK(x,y)\left[ \int^y_aK(y,z)\psi (z) dz\right]\, dy\\
&\quad =\int^x_a\left[ \int^x_z K (x,y) K(y,z)\,dy\right] \psi
(z)\,dz.\end{split}\end{equation*} Let us define the successive 
iterated kernels of $K$ by $K_1(x,z)=K(x,z)$ and
\begin{equation}\label{A7}\tag{A7} K_{n+1} (x,z) =\int^x_z K(x,y) K_n 
(y,z)\, dy.\end{equation}
Then we can write \eqref{A5} as
\begin{equation}\label{A8}\tag{A8} \phi_2 
(x)=\phi_1(x)+\epsilon^2\int^x_a K_2 (x,z) \psi (z)\,dz,\end{equation}
and similarly
\begin{equation}\label{A9}\tag{A9} \phi_3 (x)=\phi_2 (x) +\epsilon^3 
\int^x_a K_3 (x,z)\psi (z)\,dz,\end{equation}
etc., such that in general
\begin{equation}\label{A10}\tag{A10} \phi_m (x)=\phi_{m-1} 
(x)+\epsilon^m\int^x_a K_m(x,z)\psi (z)\,dz.\end{equation}

Iterating \eqref{A10} for $m=1,2,3,\ldots ,n$ and summing the 
equations results in
\begin{equation}\label{A11}\tag{A11} \phi_n(x)=\psi (x)+\int^x_a 
\left[ \sum^n_{m=1}\epsilon^mK_m(x,z)\right] \psi (z)\,dz,\end{equation}
which can be rewritten as
\begin{equation}\label{A12}\tag{A12} \psi (x)=\phi_n(x)+\epsilon 
\int^x_a\left[ -\sum^n_{m=1}\epsilon^{m-1}K_m(x,y)\right] \psi 
(y)\,dy.\end{equation}
It can be shown that the uniform limit as $n\to \infty$ exists 
(see~\cite{5,5a,5b}). 
Thus,
we obtain  equation~\eqref{A2} with
\begin{equation}\label{A13}\tag{A13} 
R(x,y)=-\sum^\infty_{n=1}\epsilon^{n-1} K_n(x,y).
\end{equation}

In case (1), $K(x,y)=k_0(y)$, the iterated kernels $K_n$ 
for $n>1$ and the resolvent kernel $R$ are in general functions of 
both $x$ and $y$.

In case (2),
$K(x,y)=k(x-y)$, i.e.\ the kernel is of the convolution (Faltung) 
type, it follows from \eqref{A7} that
\begin{equation}\label{A14}\tag{A14} 
k_{n+1}(t)=\int^t_0k(u)k_n(t-u)\,du,\end{equation}
where $x-z=t$ and $x-y=u$; therefore, all of the iterated kernels are 
of the convolution type and can be obtained by successive 
convolutions of $k$ with itself. More precisely, 
let a star denote the Faltung operation,
\begin{equation}\label{A15}\tag{A15} \phi \ast \chi (t) 
=\int^t_0\phi (t)\chi (t-u)\,du=\chi \ast \phi (t),\end{equation}
and write $\phi^{\ast\, 2} = \phi \ast \phi$, etc. Then, the resolvent 
kernel \eqref{A13} can be expressed as $R(x,y)=r(x-y)$, where
\begin{equation}\label{A16}\tag{A16} 
r(t)=-\sum^\infty_{n=1} \epsilon^{n-1}k^{\ast\, n}(t).\end{equation}

\section*{Appendix B}

In this paper, we deal with $6\times 6$ matrices of the form
\begin{equation}\label{B1}\tag{B1} \mathcal{M}=\begin{bmatrix} A & 
B\\ -B & A\end{bmatrix},\end{equation}
where $\det A\neq 0$ and $\det B=0$. The inverse of the matrix 
$\mathcal{M}$ is given by
\begin{equation}\label{B2}\tag{B2} \mathcal{M}^{-1}=\begin{bmatrix} G 
& H\\ -H & G\end{bmatrix},\end{equation}
where
\begin{equation}\label{B3}\tag{B3} G=(A+BA^{-1}B)^{-1},\quad 
H=-GBA^{-1}=-A^{-1}BG.\end{equation}

\section*{Appendix C}

Let us rewrite $P(s)$ and $Q(s)$ given by equations~\eqref{eq22} and 
\eqref{eq23} in the form
\begin{align*}\label{C1}\tag{C1} 
2\beta_0\, P(s) & = 
\frac{1-\frac{\beta_0}{w}}{1-\zeta_+}-\frac{1+\frac{\beta_0}{w}}{1+\zeta_-},\\
\label{C2}\tag{C2} 2\beta_0\, Q(s) 
&=-2+\frac{1-\frac{\beta_0}{w}}{1-\zeta_+} 
+\frac{1+\frac{\beta_0}{w}}{1+\zeta_-},\end{align*}
where $w=s\alpha$, $\beta_0=g_0\alpha$ and $\zeta_\pm$ are given by
\begin{equation}\label{C3}\tag{C3} \zeta_\pm =\frac{\beta_0}{w}\exp 
(-w\pm \beta_0).\end{equation}
If we assume that $\re (s)>g_0$, then $|\zeta_\pm |<1$. We can 
therefore expand $(1\mp \zeta_\pm)^{-1}$ in powers of $\zeta_\pm$ and 
use the relation
\begin{equation}\label{C4}\tag{C4} \mathcal{L} \left\{ 
u_{n\alpha}(t)\frac{(t-n\alpha)^{\ell-1}}{(\ell-1)!}\right\} 
=\frac{e^{-n\alpha s}}{s^\ell}\end{equation}
for integers $n\ge 0$ and $\ell >1$ to find $k(t)\to [k_1(t);k_2(t)]$. 
Here 
we use unit step functions such that $u_{n\alpha}(t)=u_0(t-n\alpha )$ 
and $u_0(t)$ is the
standard unit step function, i.e. $u_0(t)=1$ for $t\geq 0$ and 
$u_0(t)=0$ for $t<0$.

We find that $k_1=\tilde k_1J_3$,  $k_2=\tilde k_2I_3$ and
\begin{align*}\label{C5}\tag{C5} 
g^{-1}_0\tilde k_1(t)&=\mathcal{S}_1u_\alpha (t)+\mathcal{C}_2 
u_{2\alpha}(t)+\mathcal{S}_3 u_{3\alpha}
(t)+\mathcal{C}_4u_{4\alpha}(t)+\cdots ,\\
\label{C6}\tag{C6} g^{-1}_0\tilde k_2(t)+u_0(t)&=\mathcal{C}_1u_\alpha 
(t)+\mathcal{S}_2 u_{2\alpha}(t) +\mathcal{C}_3 u_{3\alpha} 
(t)+\mathcal{S}_4 u_{4\alpha}(t)+\cdots
,\end{align*}
where
\begin{equation}\label{C7}\tag{C7} \mathcal{C}_n\pm 
\mathcal{S}_n=e^{\pm n\beta_0}\left[ 
\frac{(g_0t-n\beta_0)^{n-1}}{(n-1)!}\mp 
\frac{(g_0t-n\beta_0)^n}{n!}\right]
.\end{equation}
Note that for any fixed value of $t$, only a finite number of terms 
contribute to the kernel $k(t)$.

\subsection*{Acknowledgements}

One of us (B.M.) is very grateful to Friedrich Hehl and Yuri Obukhov 
for many stimulating discussions regarding the nature of memory in
nonlocal electrodynamics.


\begin{thebibliography}{xxxx}
\bibitem{1} H.A. Lorentz, The Theory of Electrons (Dover, New York, 
1952), ch. V, \S 183, pp. 215--217.

\bibitem{2} A. Einstein, The Meaning of Relativity (Princeton 
University Press, Princeton, 1950), p. 60.

\bibitem{3} B. Mashhoon, Phys. Rev. A 47 (1993) 4498.
\bibitem{3a} B. Mashhoon, ``Nonlocal 
Electrodynamics", in: Cosmology and Gravitation, edited by M. Novello 
(Editions Fronti\`eres,
Gif-sur-Yvette, 1994), pp. 245--295. 
\bibitem{3b} U. Muench, F.W. Hehl and B. 
Mashhoon, Phys. Lett. A 271 (2000) 8.
\bibitem{3c} B. Mashhoon, ``Relativity and 
Nonlocality", in: Reference
Frames and Gravitomagnetism, edited by J.-F. Pascual-Sanchez, L. 
Floria, A. San Miguel and F. Vicente (World Scientific, Singapore, 
2001), pp. 133--144.

\bibitem{4} B. Mashhoon, Phys. Lett. A 143 (1990) 176.
\bibitem{4a} B. Mashhoon, Phys. Lett. A  145 (1990) 147.
\bibitem{4b} B. Mashhoon,
``Measurement Theory and General Relativity", in: Black Holes: 
Theory and Observation,
edited by F.W. Hehl, C. Kiefer and R. Metzler (Springer, Berlin, 
1998), pp. 269--284.

\bibitem{5} V. Volterra, Theory of Functionals and of Integral and 
Integro-Differential Equations (Dover, New York, 1959).
\bibitem{5a} H.T. Davis, 
The Theory of the Volterra
Integral Equation of Second Kind (Indiana University Studies, 17, 
1930).
\bibitem{5b} F.G. Tricomi, Integral Equations (Interscience, New York, 
1957).

\bibitem{br} N. Bohr and L. Rosenfeld, 
Det. Kgl. dansk. Vid. Selskab. 12 (1933) no. 8, translated in:
Quantum Theory and Measurement, edited by J. A. Wheeler and W. H. Zurek
(Princeton University Press, Princeton, 1983).
\bibitem{bra} N. Bohr and L. Rosenfeld, Phys. Rev. 78 (1950) 794.

\bibitem{6}  B. Mashhoon, Phys. Rev. Lett. 61 (1988) 2639.
\bibitem{6a} B. Mashhoon, Phys. Rev. Lett. 68 (1992) 3812.
\bibitem{6b} B. Mashhoon, Phys. Lett. A 198 (1995) 9.
\bibitem{6c} B. Mashhoon, Gen. Rel. Grav. 31 (1999) 681.
\bibitem{6d} B. Mashhoon, Class. Quantum Grav. 17 (2000) 2399. 
\bibitem{6e} B. Mashhoon, R. Neutze, M. Hannam and G.E. Stedman, Phys. Lett. 
A 249 (1998) 161.

\bibitem{7} L.D. Landau and E.M. Lifshitz, Electrodynamics of 
Continuous Media (Pergamon, Oxford, 1960), p. 249.
\bibitem{7a} A.C. Eringen, J. Math. Phys. 25 (1984) 3235.
\bibitem{7b}
G. Bertotti, Hysteresis in Magnetism (Academic Press, San Diego, 1998).

\bibitem{8} J.K. Roberge, in: Methods of Experimental Physics: 
Electronic Methods, vol. 2, 2nd ed., edited by E. Bleuler and R.O. 
Haxby (Academic Press, New York,
1975), ch. 12.

\bibitem{ho} F.W. Hehl and Y.N. Obukhov, in:
Gyros, Clocks, Interferometers $\cdots:$  Testing Relativistic Gravity in
Space, edited by C. L\"ammerzahl, C.W.F. Everitt and F.W. Hehl,
Lecture Notes in Physics 562 (Springer, Berlin, 2001), pp. 479--504.
\bibitem{hoa} F.W. Hehl and Y.N. Obukhov,
Foundations of Classical Electrodynamics, in press  (Birkh\"auser, Boston,
2002).

\bibitem{ll} E.H. Lieb and M. Loss, Analysis, 2nd ed.
(Amer. Math. Soc., Providence, 2001),  p. 64.

\bibitem{hen} P. Henrici, Applied and Computational Complex Analysis 
(Wiley, New York, 1986).
\end{thebibliography}
\end{document}